\documentclass{PoS}

\title{Recognition and classification of the cosmic-ray events in images captured by CMOS/CCD cameras}

\ShortTitle{Recognition and classification of the cosmic-ray events in images captured by CMOS/CCD cameras}

\author{\speaker{Michal Niedzwiecki} $^{a}$, Krzysztof Rzecki$^{a}$, Marta Marek, Piotr Homola$^{b}$, Katarzyna Smelcerz$^{ab}$,  David Alvarez Castillo$^{c}$, Karel Smolek$^{d}$, Bohdan Hnatyk$^{e}$, Jilberto Zamora-Saa$^{f}$, Alona Mozgova$^{e}$, Vahab Nazari$^{cb}$, Dariusz Gora$^{b}$, Konrad Kopanski$^{b}$, Tadeusz Wibig$^{gh}$, Alan R. Duffy$^{i}$, Jaroslaw Stasielak$^{b}$, Zoltan Zimboras$^{j}$, Marcin Kasztelan$^{k}$\\ \\ E-mail: \email{nkg@pk.edu.pl}\\
$^{a}$ Institute of Telecomputing, Faculty of Physics, Mathematics and Computer Science, Cracow University of Technology\\
$^{b}$ Institute of Nuclear Physics PAN, ul Radzikowskiego 152, 31-342 Krakow, Poland\\
$^{c}$ Joint Institute for Nuclear Research, Dubna, Russia\\
$^{d}$ Institute of Experimental and Applied Physics, Czech Technical University in Prague\\
$^{e}$ Astronomical Observatory of Taras Shevchenko National University of Kyiv\\
$^{f}$ Universidad Andres Bello, Departamento de Ciencias Fisicas, Facultad de Ciencias Exactas, Avenida Republica 498, Santiago, Chile\\
$^{g}$ University of Lodz, Faculty of Physics and Applied Informatics\\
$^{h}$ National Centre for Nuclear Physics\\
$^{i}$ Centre for Astrophysics and Supercomputing, Swinburne University of Technology, Hawthorn, VIC 3122, Australia\\
$^{j}$ Wigner Research Centre for Physics of the Hungarian Academy of Sciences\\
$^{k}$ National Centre for Nuclear Research, Andrzeja Soltana 7, 05-400 Otwock-Swierk, Poland
}

\abstract{
Muons and other ionizing radiation produced by cosmic rays and radiative decays affect CMOS/CCD sensor. When particles colliding with sensors atoms cause specific kind of noise on images recorded by cameras. We present a concept and preliminary implementation of method for recognizing those events and algorithms for image processing and their classification by machine learning. Our method consists of analyzing the shape of traces present in images recorded by a camera sensor and metadata related to an image like camera model, GPS location of camera, vertical and horizontal orientation of a camera sensor, timestamp of image acquisition, and other events recognized near-by sensors. The so created feature vectors are classified as either a muon-like event, an electron-like event or the other event, possibly noise. For muon-like events our method estimates azimuth of a muon track. Source of the data is database of CREDO (Cosmic-Ray Extremely Distributed Observatory) project and ESO  (European Southern Observatory) archives. The telescope dark frames from ESO are analysed. CREDO project collected so far over 2 millions images of events from many kinds of cameralike: smartphones camera, laptop webcams and Internet of Things cameras localised around the globe. 
}

\FullConference{36th International Cosmic Ray Conference -ICRC2019-\\
		July 24th - August 1st, 2019\\
		Madison, WI, U.S.A.}

\begin{document}

\section{Introduction}

Cosmic rays are high-energy particles from outer space. They have energy from GeV event to ZeV. When they arrive at Earth, they collide with the nuclei of atoms in the upper atmosphere, creating more particles i.a. muons. Muons do not interact strongly with matter, and can travel through the atmosphere to penetrate below ground.

Cosmic ray particles, such as muons and other radiation interact with electronics. Cosmic ray interact especially with the CCD/CMOS sensor~\cite{Groom}. When muon hits the camera sensor, it interacts with camera pixels that are on its way. As the result on image we see light dot or track depending on angle to the plane of the sensor~\cite{sensor}.

Using image analysis methods and machine learning we can detect, measure, classify and statistically analyze the cosmic-ray hits. We are presenting concept and preliminary implementation of such method.

\section{Data source}
We used two data sources. The open database of CREDO (Cosmic-Ray Extremely Distributed Observatory) project~\cite{credo} and astronomical RAW images from ESO (European Southern Observatory) archive~\cite{eso}.

CREDO project collect images which contain cosmic-ray hits from CCD/CMOS camera-based detectors. 

CREDO database contains cropped hit images in 24-bit RGB format and lossless data compression. On 2019-06-25 database contained 3139268 images from 7527 different devices: mainly from smartphones camera captured by CREDO Detector app~\cite{android}, webcam captured by application designed for Windows~\cite{windows} and RaspberryPi camera captured by application working on Raspbian OS~\cite{rpi}. Those images are ready for analysis but some of them are artifacts: hot pixels or false cosmic-ray hits from badly covered camera.  Moreover CREDO database contain hits from non CCD/CMOS based detectors (such as scintillator based detectors) but we simply ignore them. 

The hit images can be downloaded by CREDO API Tools~\cite{api}. To download those images we need user account with download privileges. 

In addition we can extract cosmic-ray hits from astronomical RAW images especially from dark frames (images from covered telescope are used for CCD noise calibration)~\cite{eso}. The ESO archive contains images in FITS format (RAW image + header). On RAW images the cosmic-ray hits must be found and cropped. 

In recent days, the world has received sensational information about a mysterious flash in the photo from the Curiosity rover on Mars. NASA say it can be cosmic-ray hit~\cite{curiosity}.

\section{Image analysis}
Cosmic-ray hit image analysis have the following steps: 1. Adjust settings and features of image source, 2. Hit detection on whole image frame and crop image with hit, 3. Extract features of hit image, 4. Classify, 5. Exclude hot pixels and artifacts and 6. Statistical analysis.

\subsection{Adjust settings and features of image source}
The source of the images is various camera sensors working on various settings and environmental conditions. The camera sensors have different pixel size and reacts differently to cosmic-ray hits.

\begin{figure}
\includegraphics[width=1\textwidth]{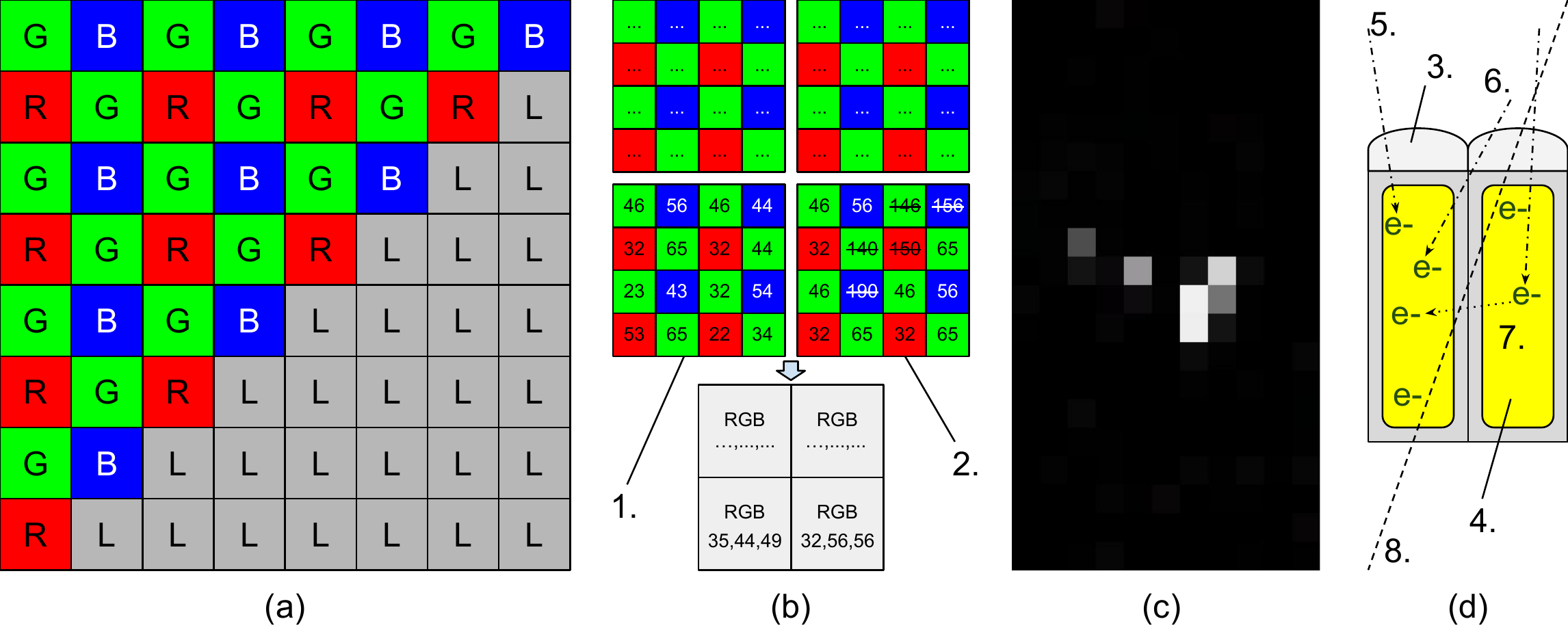}
\caption{(a) kind of color CMOS/CCD matrix vs monochrome matrix, (b) conversion pixels values from RAW image to RGB image with down-sampling (1.) and width noise reduction (2.), (c) cosmic-ray on RAW image from smartphone, (d) pixel sensor schema: optical lens (3.), photodiode (4.), expected charging by photon scoped by self lens (5.), optical crosstalk (6.), electrical crosstalk (7.), cosmic-ray (8.).  }
\label{schema}
\end{figure}

\begin{figure}
\includegraphics[width=1\textwidth]{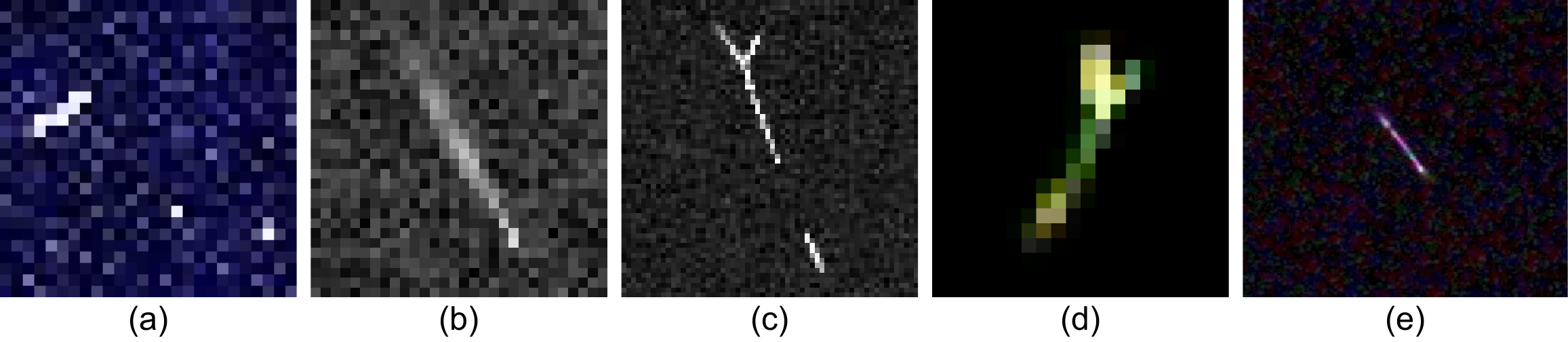}
\caption{(a) sharp cosmic ray from dark frame from ESO archive (\textasciitilde12$\mu$m LW), (b) more fuzzy cosmic ray on Sony ICX285 sensor (6.45$\mu$m LW), c) smartphone RGB camera (\textasciitilde1-2 $\mu$m LW), non-RAW image, \textasciitilde4x down-sampled image, (d) Fuzzy cosmic-ray from ESO, non-linear normalization, (e) non-continues muon track (source: DECO experiment). }
\label{cs}
\end{figure}

Cosmic ray track depend on pixel area and pixel depth sensitive  (Figure~\ref{schema}d). In science CCD/CMOS sensors pixels are big: 9-15$\mu$m length and width (LW). In smartphone sensors pixels are small: 1-2$\mu$m LW. When pixel area is large then cosmic-ray have sharp edge. On images from ESO the spot-like cosmic-ray hits may look like hot pixel (Figure~\ref{cs}a), however there are exceptions. Another ESO image have fuzzy cosmic-ray, but non-linear image-bright normalizer was used (Figure~\ref{cs}d)~\cite{Groom} and it may be electrical crosstalk effect between pixels (Figure~\ref{schema}d)~\cite{crosstalk}. In smaller pixels, the cosmic-rays are more fuzzy (Figure~\ref{cs}b), but the most samples comes from non-RAW images (Figure~\ref{cs}c). Moreover, the science CCD/CMOS camera are monochrome (Figure~\ref{schema}a). Smartphone camera is RGB where one pixel is achieved from 4 pixels for each followed color: 1x red, 1x blue and 2x green. Moreover professional (especially science) camera provide ~\cite{eso} image in RAW format, where each pixel on sensor is saved as separated pixel in image file. In smartphones without support of RAW, we get down-sampled image: each pixel in image file is the result of over an average dozen pixels in camera sensor (Figure~\ref{schema}b).

New high-end smartphones have good noise detection. It cause that detection via down-sampled images is not possible, because the single light pixels is easy to cutoff by software (Figure~\ref{schema}b). Fortunately those smartphones support saving images in RAW format where those pixels are not cutoff (Figure~\ref{schema}c).

Another environmental feature is a temperature (higher temperature, more noise, less bright distance to cosmic-ray) and GPS coordination width height and XYZ sensor orientation relative to ground plane.

\subsection{Hit detection on whole image frame and crop image with hit}
Hit detection on whole image is easier on dark frames. Each object brighter than the standard noise is a potential cosmic-ray candidate. Having series of the images of the same view of sky, we can subtract the stars. In the next phase it can be classified as hot-pixel or other artifact and cut them off.

We can use popular algorithms to detect stars, nebulae and comets on astronomical images \cite{Zheng}. In our experiment we use AstroPy~\cite{astropy} library for FITS files from ESO archive. Using AstroPy can cause some problems with non-continuous muon tracks, because AstroPy detects them separately (Figure~\ref{cs}: e and probably b, Figure~\ref{hits}d). So we must analyze detected hits by position and angle, then join them together if it looks like one hit (similar angle, are located close to each other on the extension of the line of first hit).

When the whole hit is detected we crop image around whole single cosmic-ray track. Our hit detection algorithm was implemented in~\cite{analysis}.

In CREDO database the hit detection is done using detector device app. When RAW image is not supported then typically image format is YUV with 8-bit depth of luminance (0-255 value). The detection algorithm on app is very simple and have two phases. First phase is the calibration, where dark noise is measured and bright threshold is obtained (default: 3x of average of noise but not less than 80 and not higher than 160). Second stage is the detection stage. Each image frame is analyzed and when pixel brighter than threshold is found then in the local neighborhood (in crop area range, default 60px) we find brightest pixel and crop hit image around than.

The result of this step is the set of bitmaps with cropped cosmic-ray hit (default 60px x 60px) and normalized to 0-255 bright scale. The original XY of hit coordination in original frame is stored as one of the features.

\subsection{Extract features of hit image}
The (1) and (2) step described above provides vector of features for each hit. The main features are: model of sensor (can be used to get next features e.g. pixel size, matrix area etc.), GPS location with height, image source type (RAW/non-RAW), timestamp of image, exposure time, normalization parameters and etc.

Next features are extracted by image analysis. We use photoutils~\cite{photutils} and scikit-image library~\cite{scikitimage} to extract those data. The following features are extracted: ellipticity, marked area, convex area, orientation, track length and width.

The implementation of our feature extraction algorithm is published in~\cite{hackhaton}.

\subsection{Classification}

Most papers classify hit as spot-like (dot), track-like (straight line) and worm-like (polyline) \cite{cms}\cite{Zheng}\cite{ccd}\cite{deco}\cite{crayfish}. In \cite{deco} convolutional neural networks (CNNs) was used to classify those three classes.

\begin{figure}
\includegraphics[width=1\textwidth]{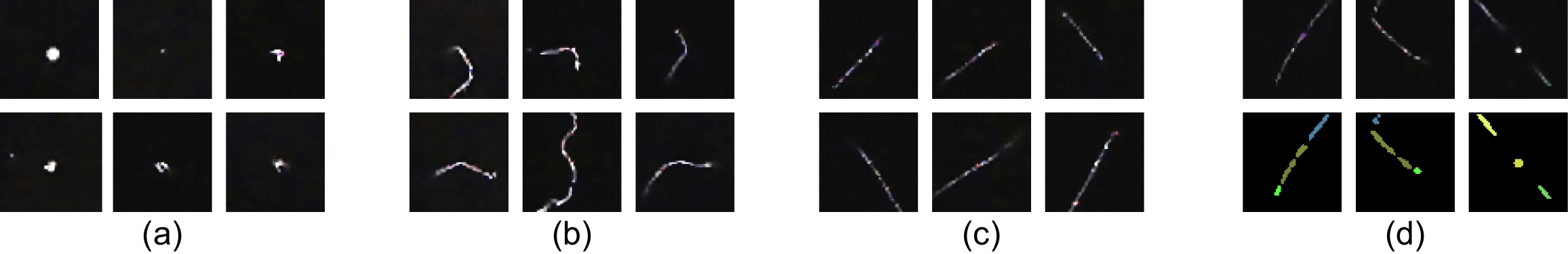}
\caption{(a) spot-like hits: ellipticity near 1, solidity near 1, (b) worm-like hits: ellipticity \textasciitilde0.5, solidity < 0.7, (c) track-like hits: ellipticity > 0.6, solidity > 0.7, (d) deblending and join track from the same hit (left), join two hits and mark third hit as separated (center), separated hits (interesting case, it is unknown whether is one hit or two separeted) (right). }
\label{hits}
\end{figure}

We can achieve the same effect by simple ellipticity and solidity (marked area per convex area) comparison (Figures~\ref{hits}a-c). 

It suppose that the tack-like hits is caused by muons hits in angle to sensor plane, worm-like is caused by $\beta$-radiation. Spot-like hits may be caused by muon or others radiation. In CORSIKA shower simulator~\cite{corsika} there is described how many kinds of particles we may expect.

\subsection{Exclude hot pixels and artifacts}

Cutting off hot pixels is easy. Having series of images, the hot pixels exist on all or on the most of frames in the same XY coordinates on image frame.

Artifact can cause a bigger problem. Most of the artifact was achieved from badly covered smartphone cameras. CREDO project uses gamification methods~\cite{gamification} for achieve new participants. In each game, we are tempted to cheat. The evidences for the artifact candidate are:
\begin{itemize}
  \item significant frequency - the frequency of cosmic-ray event about 1 per hour, bad covered camera produces hit-candidate on each frame but it will be cut off as hot-pixel, however non-standard frequency detected is evidence for artifact,
  \item non-standard hit area - especially width of track,
  \item color in RGB image.
\end{itemize}

Devices giving images with too artifacts are banned for the next analysis.

\subsection{Statistical analysis}

\begin{figure}
\includegraphics[width=1\textwidth, height=92mm]{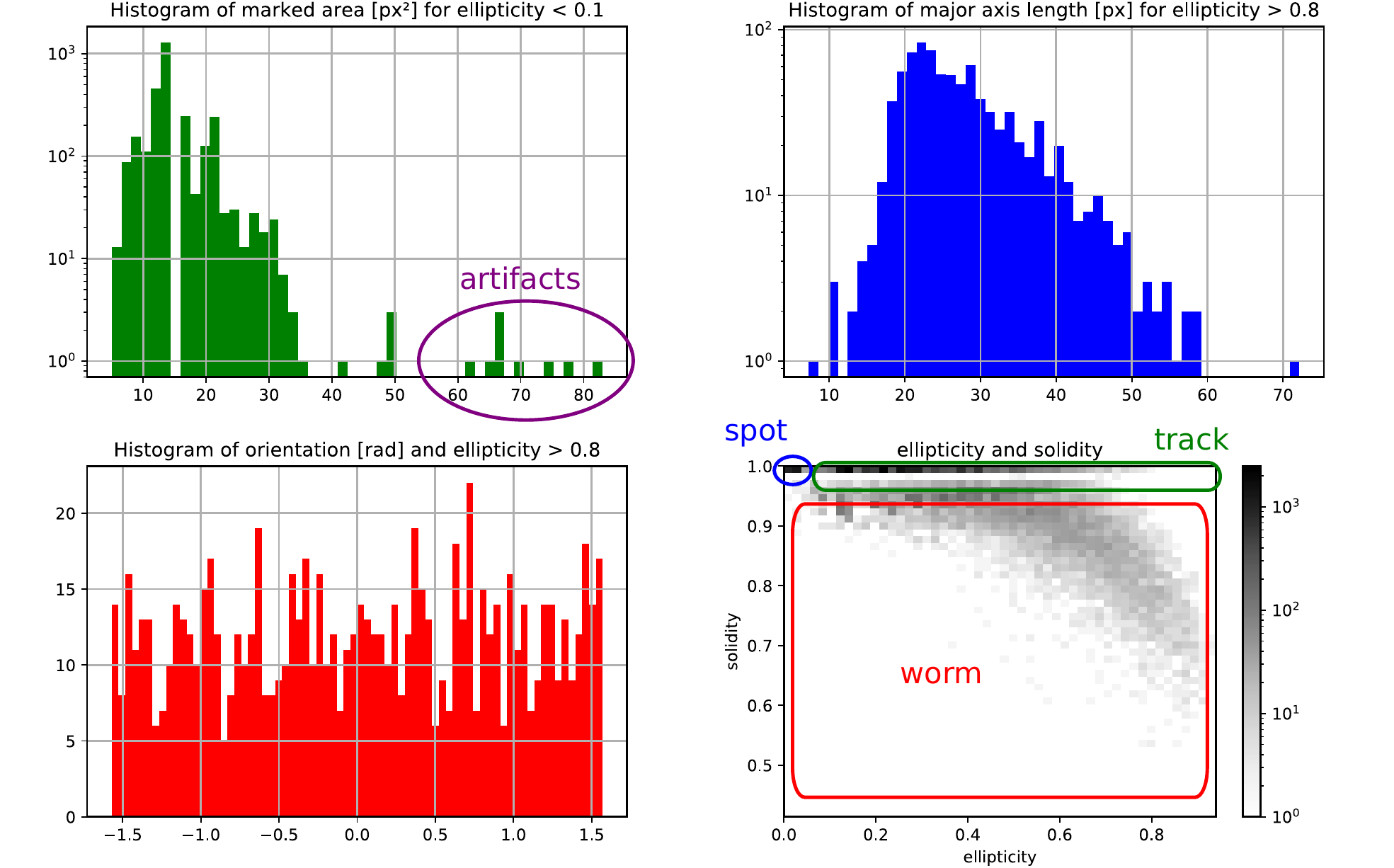}
\caption{Statistical analysis.}
\label{stats}
\end{figure}

Some statistical analysis was presented on Figure~\ref{stats} for single device from CREDO project (device ID 1397).

Top-left chart presents histogram of marked area in px$^{2}$ for hits with ellipticity value less than 0.1. Probably the hits with smaller marked area are caused by mouns and hits with larger marked area are caused by $\beta$-radiation. More experiments with physical radiation source or cosmic-ray peak simulations or statistical analysis from more devices are required in order to confirm it. Each sensor is different, so analysis must be self-calibrated to the marked area. The marked area value should depend on pixel properties (pixel size and image format: RAW or down-sampled RGB), sensitive depth and crosstalk effect. 

Top-right chart presents histogram of length of tracks in pixels for hits with ellipticity value greater than 0.8. Those hits should be caused by muons. Length of track should depend on muon angle to sensor plane, pixel properties and depth sensitive . The crosstalk effect should affect track width (minor axis length).

Bottom-left chart presents histogram of orientation of hits from top-right chart. Orientation of track will be useful, when correct azimuth is placed in hit metadata. Not every smartphone supports compass or its accuracy is \textasciitilde30$^{\circ}$. However non regular distribution of orientations on histogram, when sensor matrix is parallel to the ground flat plane, is not natural. Some noises on data transmission can produce a single track-like artefacts. Typical track-like artefact have exactly 0$^{\circ}$, 45$^{\circ}$ or 90$^{\circ}$ orientation angle.

Bottom-right chart presents 2D histogram of hit frequency with solidity and ellipticity value given. 

\section{Conclusion and future work}
Described algorithms extract various features from cosmic-ray hits on images, which can be used in future analysis.

We need to develop an auto-calibration method for each separated device, because even the same device model can provide different hit images with different frequency. Perhaps it will be possible to determine the kind of particle and measure its energy based on shape and brightness.

The detector applications for devices need to be improved by RAW image support and better GPS location acquisition. Most devices works at home where GPS signal is not found. The better time synchronization is also needed. We are going to implement NTP algorithm based on IT-SOA Service Monitoring conception~\cite{ecms}. Moreover we are going to implement some gamification ideas which will acquire new participants and increase the network of detectors. Source code of detector applications are published on MIT license so we encourage to collaboration.

It is possible to detect and analyze air showers. But current number and coverage of devices is too small. One smartphone covers approx. 1mm$^{2}$ detection area. However CREDO project adopt other non CCD/CMOS-based detectors such as CosmicWatch~\cite{cosimcwatch}, that have significantly bigger detection area.

\section{Acknowledgments}
This research has been supported in part by PLGrid Infrastructure. We warmly thank the staff at ACC Cyfronet AGH-UST, for their always helpful supercomputing support. CREDO mobile application was developed at Cracow University of Technology.


\begin{thebibliography}{99}


\bibitem{Groom}
Groom, D.,
\emph{Cosmic rays and other nonsense in astronomical CCD imagers},
\emph{Experimental Astronomy} (2002) 14: 45
[{\tt doi:10.1023/A:1026196806990}]


\bibitem{sensor}
J. Vandenbroucke et. al.,
\emph{Measurement of camera image sensor depletion thickness with cosmic rays},
\emph{Instrumentation and Detectors} 2015
[{\tt arXiv:1511.00660}]


\bibitem{credo}
The CREDO main website, accessed 16 July 2019,
[{\tt https://credo.science/credo-detector-mobile-app/}]


\bibitem{eso}
ESO Science Archive Facility, accessed 16 July 2019,
[{\tt http://archive.eso.org/}]


\bibitem{android}
CREDO Detector for Android, accessed 16 July 2019,
[{\tt https://credo.science/credo-detector-mobile-app/}]


\bibitem{windows}
CREDO Detector for Windows, accessed 16 July 2019,
[{\tt https://github.com/credo-science/CREDO-PC-Windows/}]


\bibitem{rpi}
CREDO Detector for RaspberryPi, accessed 16 July 2019,
[{\tt https://github.com/credo-science/ Credo-detector-for-linux-desktop-and-Raspberry-Pi/}]


\bibitem{curiosity}
NASA Mars Curiosity rover saw a weird light, but don't freak out,
\emph{CNET}, accessed 16 July 2019,
[{\tt https://www.cnet.com/news/ nasa-mars-curiosity-rover-saw-a-weird-light-but-dont-freak-out/}]


\bibitem{astropy}
The AstroPy project home page, accessed 16 July 2019,
[{\tt https://www.astropy.org/}]





\bibitem{photutils}
The scikit-image project home page, accessed 16 July 2019,
[{\tt https://photutils.readthedocs.io/}]


\bibitem{scikitimage}
The scikit-image project home page, accessed 16 July 2019,
[{\tt https://scikit-image.org/}]


\bibitem{cms}
CMS Collaboration,
\emph{Measurement of the charge ratio of atmospheric muons with the CMS detector},
\emph{Phys.Lett.B}692:83-104,2010
[{\tt arXiv:1005.5332}]

\bibitem{Zheng}
Zheng, C. X.; Pulido, J.; Thorman, P.; Hamann, B.,
\emph{An improved method for object detection in astronomical images},
\emph{MNRAS} 451 (4):4445-4459. 2015.
[{\tt doi:10.1093/mnras/stv1237}]


\bibitem{ccd}
M. Fisher-Levine and A. Nomerotski,
\emph{Characterising CCDs with cosmic rays},
\emph{JINST} (2015)
[{\tt doi:10.1088/1748-0221/10/08/c08006}]


\bibitem{hackhaton}
Niedzwiecki, M.,
\emph{ESO archive cosmix-ray extraction script}, accessed 16 July 2019,
[{\tt https://github.com/credo-science/credo-hackhaton-nkg/}]


\bibitem{analysis}
Niedzwiecki, M.,
\emph{scripts for anaysis images from CREDO project database}, accessed 16 July 2019,
[{\tt https://github.com/dzwiedziu-nkg/credo-analysis/}]


\bibitem{api}
ACC Cyfronet AGH-UST,
\emph{API tools for CREOD project}, accessed 16 July 2019,
[{\tt https://github.com/credo-science/credo-api-tools/}]


\bibitem{deco}
Matthew Meehan et al.,
\emph{The particle detector in your pocket: The Distributed Electronic Cosmic-ray Observatory},
\emph{Instrumentation and Methods for Astrophysics} ICRC 2017
[{\tt arXiv:1708.01281}]


\bibitem{crayfish}
The CRAYFISH main website, accessed 16 July 2019,
[{\tt https://crayfis.io/}]


\bibitem{corsika}
D. Heck and T. Pierog,
\emph{Extensive Air Shower Simulation with CORSIKA: A User’s Guide},
\emph{KIT - Universitat des Landes Baden-W ¨ urttemberg und nationales Forschungszentrum in der Helmholtz-Gemeinschaft}, accessed 16 July 2019,
[{\tt https://web.ikp.kit.edu/corsika/usersguide/usersguide.pdf}]


\bibitem{gamification}
Huotari, K., Hamari, J.,
\emph{Defining Gamification – A Service Marketing Perspective},
\emph{Proceedings of the 16th International Academic MindTrek Conference 2012} Tampere, Finland, October 3-5
[{\tt doi:10.1145/2393132.2393137}]


\bibitem{ecms}
M. Niedzwieck et.al.,
\emph{Asynchronous Agent System For Monitoring Communication And System States Based On The SOA Paradigm},
\emph{25th Conference on Modelling and Simulation}, 2011
[{\tt ISBN: 978-0-9564944-2-9}]


\bibitem{cosimcwatch}
The CosmicWatch project main website, accessed 16 July 2019,
[{\tt http://www.cosmicwatch.lns.mit.edu/}]


\bibitem{pixel}
Sungsoo Choi, KyungHo Lee, et. al.,
\emph{An all pixel PDAF CMOS image sensor with 0.64μmx1.28μm photodiode separated by self-aligned in-pixel deep trench isolation for high AF performance}
\emph{Symposium on VLSI Technology} 2017
[{\tt doi:10.23919/VLSIT.2017.7998212}]

\bibitem{crosstalk}
C. R. Zhao et al.,
\emph{Adaptive Pixel Crosstalk Compensation for CMOS Image Sensor}
\emph{Advanced Materials Research } Vol. 981, pp. 310-314, 2014
[{\tt doi:10.4028/www.scientific.net/AMR.981.310}]

\end{thebibliography}
\end{document}